\documentclass[showpacs, showkeys,twocolumn,preprintnumbers,floatfix,amsmath,amssymb]{revtex4}
\usepackage{natbib}
\usepackage{amsmath,amsthm}
\usepackage{float}
\usepackage{graphicx}
\usepackage{caption}
\usepackage{subfigure}
\usepackage{tabularx}
\usepackage{threeparttable}
\usepackage{dcolumn}
\usepackage{bm}
\usepackage{color,epsfig,multirow}
\usepackage{array}
\usepackage{hyperref}
\usepackage{algorithm}
\usepackage{algorithmic}

\theoremstyle{definition}

\theoremstyle{remark}

\allowdisplaybreaks[4]

\begin{document}
\preprint{Preprint}

\title{Harmonic surface mapping algorithm for fast electrostatic sums}
\author{Qiyuan Zhao}\affiliation{Zhiyuan College, Shanghai Jiao Tong University, Shanghai 200240, China}
\author{Jiuyang Liang}\affiliation{School of Mathematical Sciences, Shanghai Jiao Tong University, Shanghai 200240, China}
\author{Zhenli Xu}\email{xuzl@sjtu.edu.cn}
\affiliation{School of Mathematical Sciences, Institute of Natural Sciences and MoE Key Lab of \\Scientific and Engineering Computing, Shanghai Jiao Tong University, Shanghai 200240, China}
\date{\today}
\begin{abstract}
We propose a harmonic surface mapping algorithm (HSMA) for electrostatic pairwise sums of an infinite number of image charges.
The images are induced by point sources within a box due to a specific boundary condition which can be non-periodic.
The HSMA first introduces an auxiliary surface such that the contribution of images outside the surface can be approximated by
the least-squares method using spherical harmonics as basis functions. The so-called harmonic surface mapping is the procedure
to transform the approximate solution into a surface charge and a surface dipole over the auxiliary surface, which becomes point images
by using numerical integration. The mapping procedure is independent of the number of the sources and is considered to have
a low complexity.  The electrostatic interactions are then among those charges within the surface
and at the integration points, which are all the form of Coulomb potential and can be accelerated straightforwardly
by the fast multipole method to achieve linear scaling. Numerical calculations of the Madelung constant of a crystalline lattice,
electrostatic energy of ions in a metallic cavity, and the time performance for large-scale systems show that
the HSMA is accurate and fast, and thus is attractive for many applications.

\end{abstract}



\maketitle

\section{Introduction}

Calculations of electrostatic interactions are among the most important components in molecular simulations
of systems at the micro/nano scale such as biomolecules, membranes, electrochemical energy devices,
and soft materials \cite{frenkel2002book,FPP+:RMP:2010,WK+:N:2011}. Due to the long-range nature, fast algorithms have to be used to speed up the pairwise
interactions within a finite simulation volume specified a boundary condition. To mimic the system environment,
a periodic boundary condition (PBC) is often used, and mostly, Ewald-based lattice summations such
particle mesh Ewald \cite{DYP:JCP:1993,EPB+:JCP:1995,Peterson:JCP:1995} or
particle-particle particle-mesh method \cite{HE::1988,SKT:JCC:1993,LDT+:MS:1994} have been employed
for the electrostatic interaction. When the simulation systems are nonperiodic
such as those with interfaces, the 3D Ewald-based techniques can be time-consuming since a larger simulation volume has to
be used to reduce the artifact due to the PBC assumption, and the development of new techniques such as
Ewald-type methods for quasi-2D systems \cite{Mazars:PR:2011,ArnoldHolm:CPC:2002,ArnoldDH:JCP:2002,Hu:JCTC:2014,dosSantosGL:JCP:17}
and non-Ewald methods \cite{FN:BR:2012,CDJ:JCP:2007,LXX:NJP:2015} remains the crucial theme for simulating electrostatic phenomena.

In many problems, electrostatic interactions in a box with specific boundary conditions can be represented
as the sum of an infinite number of charged particles using the method of images and recursive reflections between
boundary faces \cite{Jackson::2001}. The objective of this paper is to develop
a harmonic surface mapping algorithm (HSMA) to transform the infinite sum into a finite one such that the accelerating techniques
such as the fast multipole method (FMM) \cite{GR:JCP:1987,GR:AN:1997,YBZ:JCP:2004,CGR:JCP:1999} can be simply applied to
achieve an $O(N)$ complexity. The HSMA removes the difficulty of solving boundary integral equations
by introducing an auxiliary surface away from the central box. The idea of the
auxiliary surface has been used in scattering problems \cite{GillmanBarnett:JCP:2013}, multiphase flows \cite{MarpleBarnett:SISC:2016},
and electrostatics \cite{gumerov2014method,ZhaoLiuXu:CCP:2018}, which is successful
because it allows that the nearest-neighbor interactions within the surface is summed directly and the distant interactions can
be approximated by a small number of basis functions using least-squares fittings. The harmonic surface mapping developed in this work maps the
contribution of the distant interactions into a surface integral such that it can be approximated by discrete
images on surfaces with a high order of convergence due to the use of the Fibonacci integration.
Essentially, The HSMA can belong to the method of fundamental solutions \cite{shigeta2012adaptive,ChoBarnett:PE:2015}, and dramatically
all approximate fundamental functions (the images) of the HSMA are located on the auxiliary surfaces and their strengths
are obtained by the harmonic surface mapping and the numerical integration, and thus the convergence can be then ensured. Numerical examples
are performed to show the efficiency of the HSMA which is accelerated by both the FMM and graphics processors.

The HSMA is potentially useful for many electrostatic problems in presence of non-periodic boundaries.
For instance, for quasi-2D systems with ions confined by parallel dielectric/metallic media, the images within
the auxiliary surface are produced by a combination of the mirror reflections and the periodic extension,
for which the direct summation of the infinite reflective images can be costly. Moreover, the HSMA can
be useful for systems with irregular boundary once the images within the auxiliary surface can be well
approximated. We show an example to demonstrate the algorithm can accurately calculate the electrostatic interaction
of ions within metallic boundaries, which is considered difficult due to the divergence of the traditional image-reflection method.

\section{Method}

Consider a charged system of $N$ point sources located at $\{\mathbf{r}_j, j=1,\cdots, N\}$ in a cubic domain $\Omega$ of side length $L$.
Let $\Phi(\mathbf{r})$ be the electric potential distribution due to these point sources and a specific boundary condition.
Within $\Omega$, the potential satisfies the Poisson's equation,
\begin{equation}
-\nabla^2\Phi(\mathbf{r}) =4\pi\sum_{j=1}^N q_j \delta(\mathbf{r}-\mathbf{r}_j),
\end{equation}
and in many situations, the solution can be expressed as an infinite sum,
\begin{equation}
\Phi(\mathbf{r}) =  \sum_{j=1}^N \frac{q_j}{|\mathbf{r}_j-\mathbf{r}|} +\sum_{i=N+1}^\infty
\frac{q_i}{|\mathbf{r}_i-\mathbf{r}|}, \label{sum}
\end{equation}
where the first term is the direct potential of the point sources and the second term describes the contribution of the infinite number of image charges
which are introduced to satisfy the boundary condition. For instance, if the boundary condition is periodic,
one has,
\begin{equation}
\Phi(\mathbf{r}) = \sum_\mathbf{k}  \sum_{j=1}^N \frac{q_j}{|\mathbf{r}_j+\mathbf{k}L-\mathbf{r}|}
\end{equation}
where $\mathbf{k}$ runs over all three-dimensional integer vectors and the images are periodic copies
of the source charges; if the boundary is a dielectric interface, i.e.,
the boundary conditions become the continuities of the potential and electric displacement, then one gets \cite{YLL:JPCB:2002},
\begin{equation}
\Phi(\mathbf{r}) = \sum_\mathbf{k}\sum_{j=1}^N \frac{(\gamma)^{|\mathbf{k}|}q_j}{|\mathbf{r}_{j,\mathbf{k}}+\mathbf{k}L-\mathbf{r}|} \label{diel}
\end{equation}
where $\gamma=(1-\varepsilon)/(1+\varepsilon)$ with $\varepsilon$ being the dielectric ratio between
the exterior and the interior of the simulation box, $|\mathbf{k}|$ represents the sum
of absolute values of all components,
and $\mathbf{r}_{j,\mathbf{k}}=(-1)^\mathbf{k}\circ \mathbf{r}_j$ uses the element-by-element power and product,
and the images are constructed by iterative reflections of the sources due to the six boundary faces.
Similar sums can be found when the boundary condition is the combination of different types of boundary conditions
such as periodic, Dirichlet, Neumann, Robin and dielectric-jump conditions.
These infinite sums are often slowly convergent (dielectric boundaries), conditionally convergent (periodic boundaries)
or even divergent (conducting limit $\gamma\rightarrow-1$ of dielectric boundaries), and the direct pairwise summation
for the energy and force calculation is often difficult. In practical simulations, Ewald-based lattice-summation
methods \cite{DYP:JCP:1993,EPB+:JCP:1995} are often employed for an $O(N\log N)$ calculations through the use of FFT acceleration. These methods are
efficient for periodic systems, but less efficient for nonperiodic systems considering a large buffer zone has to
be introduced to remove the artifact of periodicity. Alternatively, non-Ewald methods such as reaction-field methods \cite{KW:JCP:1989,AL:JCP:1993,LBD+:JCP:2009,BR:JCP:1994}
have been studied in literature for calculating the long-range interaction.
For the Poisson's equation or the modified Helmholtz equation in 2D, the method of images has been
used to solve the boundary-value problem such that the adaptive FMM can be used to calculate the volume integral \cite{EthridgeGreengard:SISC:2001, ChengHL:JCP:2006}.

In this work, we introduce the HSMA for the summation of infinite charges. It is composed of two steps: one approximates the infinite sum outside a
given surface by a harmonic series, followed by a harmonic mapping from this series onto a surface integral which
can be discretized into contributions of point charges and thus the FMM or GPU acceleration can be directly
applied. We describe the details of the HSMA in the following content.

\begin{figure}[htbp]
		\centering
		\includegraphics[width=0.8\linewidth]{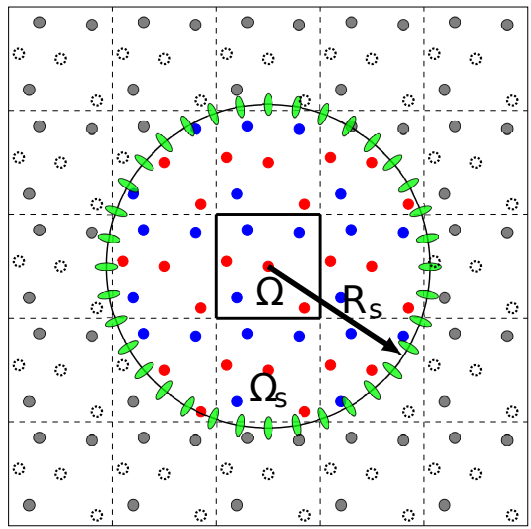}
		\caption{Schematic illustration of the HSMA for an infinite particle system, where $\Omega$ is the
central box and particles within it are the sources. $\partial \Omega_\mathrm{s}$ is an auxiliary boundary.
The contribution of infinite images outside $\partial\Omega_\mathrm{s}$ is approximated by point and dipolar images (elliptical symbols) on the surface.}
		\label{schem}
	\end{figure}

\subsection{Spherical harmonics approximation}
Consider the pairwise sum of the infinite particle system given in Eq. \eqref{sum}. As schematically shown in Fig. \ref{schem},
we introduce an artificial boundary $\partial\Omega_\mathrm{s}$ of spherical shape to separate
the infinite system such that domain $\Omega_\mathrm{s}$ includes $\Omega$ and they share the same center at the origin. The spherical radius
is $R_\mathrm{s}$. Let $\{q_j, j=1,\cdots, N_\mathrm{s}\}$ be the set of those charges within the sphere,
where the first $N$ charges are the sources and the remaining $(N_\mathrm{s}-N)$ charges are the images near the central box.
The electric potential can be then rewritten into the sum of two contributions,
$\Phi(\mathbf{r}) = \Phi_\mathrm{in} +\Phi_\mathrm{out},$
where
\begin{equation}
\Phi_\mathrm{in}=\sum_{j=1}^{N_\mathrm{s}} \frac{q_j}{|\mathbf{r}_j-\mathbf{r}|},
\end{equation}
and $\Phi_\mathrm{out}$ is the electric potential produced by all images outside $\partial\Omega_\mathrm{s}$.

Since $\Phi_\mathrm{out}$ is a harmonic function in
domain $\Omega$, it can be approximated by a truncated spherical harmonic series \cite{gumerov2014method}
using the spherical coordinates,
\begin{equation}
	\Phi_\mathrm{out}= \sum_{n=0}^{P}\sum_{m=-n}^{n}A_{nm} B_n^m(r,\vartheta,\varphi), \label{harmonic}
\end{equation}
where $B_n^m=r^nY_n^m(\vartheta,\varphi)$ are orthogonal bases and
$Y_n^m$ is the spherical harmonic function of degree $n$ and order $m.$  The number of
the basis functions is $N_\mathrm{b}=(P+1)^2$.

To determine
the coefficients $\{A_{nm}\}$, one shall select $N_\mathrm{m}$ monitoring points, $\{\mathbf{x}_i, i=1,\cdots, N_\mathrm{m}\}$,
which are nearly uniformly distributed on the boundary, and
find the coefficients such that the boundary condition (nonperiodic) is satisfied at these points.
Let $\mathcal{L}(\Phi,\partial_\mathbf{n}\Phi)=0$ for $\mathbf{r}\in\partial\Omega$ be the boundary condition for the Poisson's equation,
where $\mathcal{L}$ is a linear operator and $\Phi$ is the sum of $\Phi_\mathrm{in}$ and $\Phi_\mathrm{out}$.
The $N_\mathrm{m}$-dimensional residual vector $\mathbf{v}=(v_1,\cdots,v_{N_\mathrm{m}})^T$ is then defined by these monitoring points
such that
$v_i =  \mathcal{L}(\Phi(\mathbf{x}_i),\partial_\mathbf{n}\Phi(\mathbf{x}_i)).$
The coefficients of the truncated spherical harmonic series are calculated by minimizing the $L_2$ norm
of the residual vector,
\begin{equation}\{A_{nm}\}=\mathrm{argmin} \|\mathbf{v}\|_2.\end{equation}
Because the number of monitoring points is usually bigger than the number of basis functions, $N_\mathrm{m}>N_\mathrm{b}$,
the minimization is generally done by using the discrete least-squares method, resulting
in an $N_\mathrm{b}\times N_\mathrm{b}$ linear system. The contribution of $\Phi_\mathrm{in}$
appears in the right hand side of the system and the complexity to calculate it is
$O(N_\mathrm{s}N_\mathrm{m})$ without the acceleration or $O(N_\mathrm{s}+N_\mathrm{m})$
with the FMM acceleration.

The PBC is different from other boundary conditions, as the distribution of
monitoring points on $\partial\Omega$ would lead to an ill-conditioned fitting matrix.
To avoid this problem, one should choose the monitoring points $\{\mathbf{x}_i\}$ on the circumsphere of the central box
$\Omega$. Let $\widetilde{\mathbf{x}}_i=\mathbf{x}_i-\mathbf{k} L$ with $\mathbf{k}$ being
the index vector such that $\widetilde{\mathbf{x}}_i$ is located within the central box.
Then the $i$th component of the residual vector is then given by $v_i=\Phi(\mathbf{x}_i)-\Phi(\widetilde{\mathbf{x}}_i)$.
The minimization for the $L_2$ norm of the residual vector is then better conditioned.
The condition number can be further reduced when the monitoring points are uniformly distributed over the circumsphere.
In this work, we use the Fibonacci grid \cite{swinbank2006fibonacci}, which arranges the monitoring points
along a spiral lattice such that they are close to a uniform distribution on the spherical surface,
namely, the polar and azimuth angles of these points are $\vartheta_i=\arcsin((2i-1)/N_\mathrm{m}-1)$ and
$\varphi_i=2i\pi \omega$ for $i=1,\cdots, N_\mathrm{m}$, with $\omega=(\sqrt{5}-1)/2$ being the golden ratio.

Let $\alpha$ be the ratio between $R_\mathrm{s}$ and the
radius of the circumsphere of the central box. An error bound of the spherical
harmonic expansion for the electric potential truncated at $n=P$ is given by
\cite{gumerov2014method},
\begin{equation}
\epsilon_\mathrm{potl}\leq\frac{4\pi d_0 R_\mathrm{s}^2 \max|q_i|}{(\alpha-1)(P-2)}\left(\frac{1}{\alpha}\right)^{P},
\end{equation}
where $d_0$ is a constant which is close to the number density of the particles.
Since the force on each particle is used in molecular dynamics simulations, it is important to estimate the error of
the gradient of the potential. Let $\epsilon_\mathrm{grad}^\ell$ be the $\ell$th component
of the three-dimensional error vector, then the error bound of the potential gradient is,
\begin{equation}
\epsilon_\mathrm{grad}^\ell\leq\frac{4\pi d_0R_s\max|q_i|}{(\alpha-1)(P-2)}\left(P+\frac{\alpha}{\alpha-1}\right)\left(\frac{1}{\alpha}\right)^{P-1},
\end{equation}
for $\ell=1,2,3$. See Appendix for the proof.
Due to the spectral convergence with the degree of the harmonics, it is expected that a small value of $P$
can provide an accurate approximation of the solution.

\subsection{Harmonic surface mapping}

We use the Green's second identity to transform the spherical harmonic expansion of $\Phi_\mathrm{out}$ into a surface integral over $\partial\Omega_\mathrm{s}$.
This will further speed up the calculation because the approximation to the surface integral leads to
the sum of point images, and state-of-the-art accelerating techniques can be simply used. Essentially, the so-called harmonic surface mapping
is the relation between the spherical harmonic bases and the fundamental solution bases. It avoids the solution
of ill-conditioned linear systems by directly using the images on surfaces for the least-squares fitting since it benefits
from the orthogonality of spherical harmonics.

Let $G(\mathbf{r},\mathbf{r}')=1/|\mathbf{r}-\mathbf{r}'|$ be the free-space Green's function.
Since $\Phi_\mathrm{out}$ satisfies the Laplace equation $\nabla^2\Phi_\mathrm{out}=0$ for $\mathbf{r}\in\Omega$,
using the Green's second identity leads us to,
\begin{equation}
\Phi_\mathrm{out} = \frac{1}{4\pi} \int_{\partial \Omega_\mathrm{s}} \left[
 G \frac{\partial \Phi_\mathrm{out}(\mathbf{r}')}{\partial \mathbf{n}'}
- \Phi_\mathrm{out}(\mathbf{r}') \frac{\partial G }{\partial \mathbf{n}'} \right] dS', \label{green}
\end{equation}
where $\mathbf{n}'$ represents the unit outer normal direction at $\mathbf{r}'$.
Eq. \eqref{green} describes that the potential $\Phi_\mathrm{out}$ is a sum of
a surface charge and a surface dipole, which are analytically given by the spherical harmonic series.
Defining
$\sigma(\mathbf{r}')=\partial_{\mathbf{n}'} \Phi_\mathrm{out}(\mathbf{r}')$ and using Eq. \eqref{harmonic},
one has,
\begin{equation}
\sigma(\mathbf{r}')=\frac{1}{r'}\sum_{n=0}^{P}\sum_{m=-n}^{n} n A_n^m B_n^m(r',\vartheta',\varphi').
\end{equation}
In order to calculate the surface dipole in Eq. \eqref{green} in the way of the Coulomb form and
thus the HSMA is more efficient, the surface dipole can be approximated by two surface charges. The normal derivative of the Green's function in Eq. \eqref{green}
is approximated by the central difference,
\begin{equation}
\frac{\partial G }{\partial \mathbf{n}'}=\frac{1}{\Delta r}\left[
G(\mathbf{r}^+,\mathbf{r})-G(\mathbf{r}^-,\mathbf{r})\right], \label{cd}
\end{equation}
with $\mathbf{r}^\pm=(r'\pm\Delta r/2,\vartheta', \varphi').$ Let $\Omega_\mathrm{s}^\pm$
represent the spheres of radii $R_\mathrm{s}\pm\Delta r/2$, which are concentric with $\Omega_\mathrm{s},$
and let us define
\begin{equation}\mu^\pm(\mathbf{r}')=\frac{R_\mathrm{s}^2}{\Delta r(R_\mathrm{s}\pm\Delta r/2)^2}\Phi_\mathrm{out}(\mathbf{r}').
\end{equation}
We can then write $\Phi_\mathrm{out}$ as the sum of the following three integrals:
\begin{equation}\begin{aligned}
\Phi_\mathrm{out}(\mathbf{r})=\frac{1}{4\pi}\left(\int_{\partial\Omega_\mathrm{s}}\frac{\sigma(\mathbf{r}')}{|\mathbf{r}-\mathbf{r}'|}dS'
+\int_{\partial\Omega_\mathrm{s}^-}\frac{\mu^-(\mathbf{r}')}{|\mathbf{r}-\mathbf{r}'|}dS' \right.\\
\left.-\int_{\partial\Omega_\mathrm{s}^+}\frac{\mu^+(\mathbf{r}')}{|\mathbf{r}-\mathbf{r}'|}dS'
\right),~~~~~~~
\end{aligned} \label{integrals}
\end{equation}
where the charge densities on the three surfaces are given analytically. The last two terms are from
the approximation of the surface dipole, which is an $O(\Delta r^2)$ approximation and thus is accurate
when $\Delta r$ is small.

The surface integrals are not singular and can be approximated by traditional numerical quadratures.
One difficulty for integral over a spherical surface is the appropriate distribution of sampling points as the
crystal grids will include defects leading to the loss of accuracy if uniform weights are applied. Fibonacci numerical integration \cite{hannay2004fibonacci} is introduced
to optimize the approximation of the integrals, which achieves the order of accuracy $\sim N_\mathrm{o}^{-6}$
for $N_\mathrm{o}$ grid points. Let $F_1$ and $F_2$ be two successive Fibonacci numbers with $F_1<F_2$,
and $f(\mathbf{r}')$ be the integrand. The Fibonacci numerical integral is written as,
\begin{equation}
\int_{\partial \Omega_\mathrm{s}} f(\mathbf{r}') dS'\approx \frac{2\pi R_\mathrm{s}^2}{F_2} \sum_{j=0}^{F_2}[1+\cos(\pi z_j)][f(\mathbf{r}_{2j+1})+f(\mathbf{r}_{2j+2})],
\end{equation}
where $z_j=(-1+2j/F_2)$, $\mathbf{r}_{2j+1}=(R_\mathrm{s},\vartheta_j, \varphi_j)$, $\mathbf{r}_{2j+2}=(R_\mathrm{s},\vartheta_j, \pi+\varphi_j)$,
$\vartheta_j=\arccos(z_j+\sin(\pi z_j)/\pi)$ and $\varphi_j=\pi jF_1/F_2.$ Using this quadrature to the three integrals
in Eq. \eqref{integrals}, we transform the contribution of infinite images outside $\partial\Omega_\mathrm{s}$
into finite images on $\Omega_\mathrm{s}$ and $\Omega_\mathrm{s}^\pm$,
\begin{equation}
\Phi_\mathrm{out}(\mathbf{r})=\sum_{j=1}^{N_\mathrm{o}} \frac{Q_j}{|\mathbf{R}_j-\mathbf{r}|},
\end{equation}
where $\mathbf{Q}_j$ and $\mathbf{R}_j$ are the charge and location of the $j$th one of a total of
$N_\mathrm{o}=6(F_2+1)$ images.

The numerical integrations over the surfaces are accurate as long as the integration points are not small due to the
sixth order of convergence, and thus the dominant error for the mapping is due to the finite-difference approximation of
the surface dipole. The truncation error for the approximation given in Eq. \eqref{cd} can be expressed as
$\Delta r^2 \partial^3_{r'}G(\mathbf{r},\mathbf{r}')/24$ for $r'\in[R_\mathrm{s}-\Delta r/2,R_\mathrm{s}+\Delta r/2]$,
which has a small prefactor of $\Delta r^2$ in case that $\alpha$ is not close to one.

\subsection{Algorithm steps and complexity}

We describe the details of the HSMA which is composed of the preparation and update steps.

The preparation step generates monitoring
points $\{\mathbf{x}_i, i=1,\cdots, N_\mathrm{m}\}$ and integration points $\{\mathbf{R}_j, j=1,\cdots, N_\mathrm{o}\}$,
and constructs the fitting and mapping matrices.
The monitoring points are distributed either on the circumsphere of the central box $\Omega$ for a periodic boundary condition
or on $\partial\Omega$ for other boundary conditions. The fitting matrix
does not depend on the source charges, and the QR factorization for the least squares is performed in this step. The mapping matrix
calculates the spherical harmonic expansion at the integration points and its $(\ell, k)$ entry represents the value of the $k$th harmonic basis
function at point $\mathbf{R}_\ell$. In the calculation, the complex basis functions $B_n^{\pm|m|}(\mathbf{r})$ can be redefined
into two real functions with the real and imaginary parts and a recursive process is introduced to calculate these functions.

In the update step, images within $\Omega_\mathbf{s}$ are first generated. The least-square problem is solved
to determine the coefficients $\{A_{nm}, n=0,\cdots, P ~\hbox{and}~ m=-n,\cdots, n\}$ of the spherical harmonic expansion,
followed by the calculation of the charges of the images $\{Q_j, j=0,\cdots, N_\mathrm{o}\}$ at the integration points.
The electric energy and forces of each source can be determined by summing up the contribution from both the charges
$\{q_j, j=1,\cdots, N_\mathrm{s}\}$ and $\{Q_j, j=1,\cdots, N_\mathrm{o}\}$.

Since the number of sources $N$ is generally much bigger than the number of monitoring points $N_\mathrm{m}$,
the number of bases $N_\mathrm{b}$ and the number of integration points $N_\mathrm{o}$, the most time-consuming
calculations are the determination of the potential due to the $N_\mathrm{s}$ charges within sphere $\Omega_\mathrm{s}$.
Both FMM and GPU accelerations are applied to speed up the calculations. The complexity of each step of the algorithm
with the FMM acceleration is summarized in Table \ref{complexity}. By using the FMM, the complexity of calculating
the right hand side of the linear least-squares system is reduced to $O(N_\mathrm{s} + N_\mathrm{m})$, and the solution
of the system is $O(N_\mathrm{b}N_\mathrm{m})$ as the QR factorization has been done in the preparation step. Moreover,
the use of FMM reduces the complexity of the force and energy calculation to $O(N_\mathrm{s}+N_\mathrm{o})$. The GPU
acceleration, which has been widely used in different electrostatics
algorithms \cite{BrownKPT:CPC:2012,LeGrandGW:CPC:2013,GengJ:CPC:2013,AdelmanGD:IEEETM:2017},
is also implemented for our computer program. It does not reduce the asymptotic scaling,
but it is usually much faster than the single-core FMM when the number of sources is no more than one million.

\begin{table}[htbp]
		\caption{Complexity of the HSMA using the FMM}
		\label{complexity}
		\begin{tabular}{p{5cm}<{\centering}  p{3cm}<{\centering}}
			\hline
			Preparation step & Complexity \\\hline
			Construct the fitting matrix & $O(N_\mathrm{m}N_\mathrm{b})$\\
			Factorize the fitting matrix& $O(N_\mathrm{b}^3)$\\
			Construct the mapping matrix & $O(N_\mathrm{o}N_\mathrm{b})$\\\hline\hline
			Update step & Complexity \\\hline
			Generate images in $\Omega_\mathbf{s}$ & $O(N_\mathrm{s})$\\
			Solve the least-squares problems & $O(N_\mathrm{s} + N_\mathrm{m} + N_\mathrm{b}N_\mathrm{m})$\\
			Assign images on surfaces & $O(N_\mathrm{b}N_\mathrm{o})$\\
			Calculate the forces and energies   & $O(N_\mathrm{s}+N_\mathrm{o})$\\
			\hline
		\end{tabular}
	\end{table}

\section{Results}

\begin{figure*}[t!]
	\centering
	\includegraphics[width=0.49\linewidth]{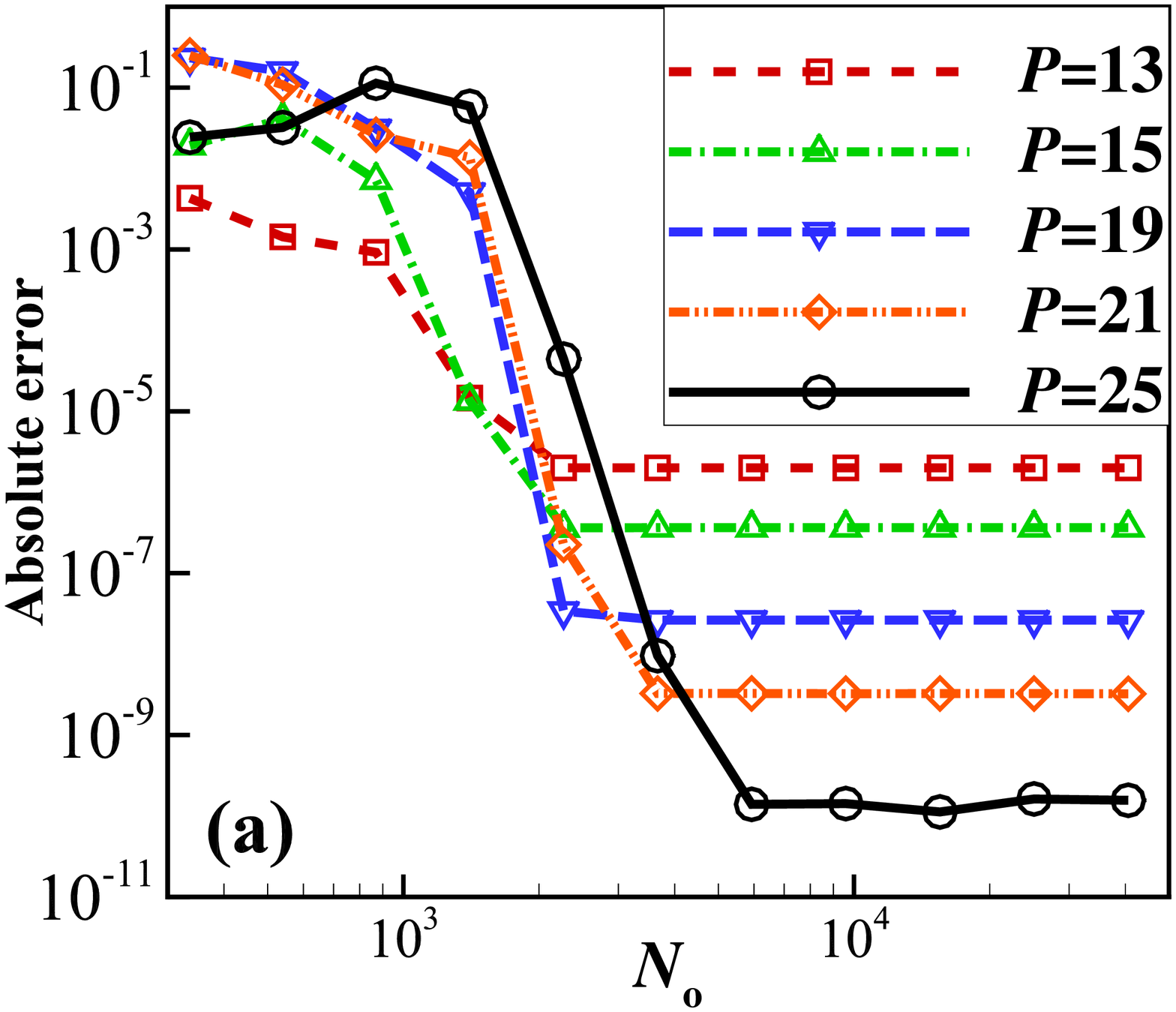}
	\includegraphics[width=0.49\linewidth]{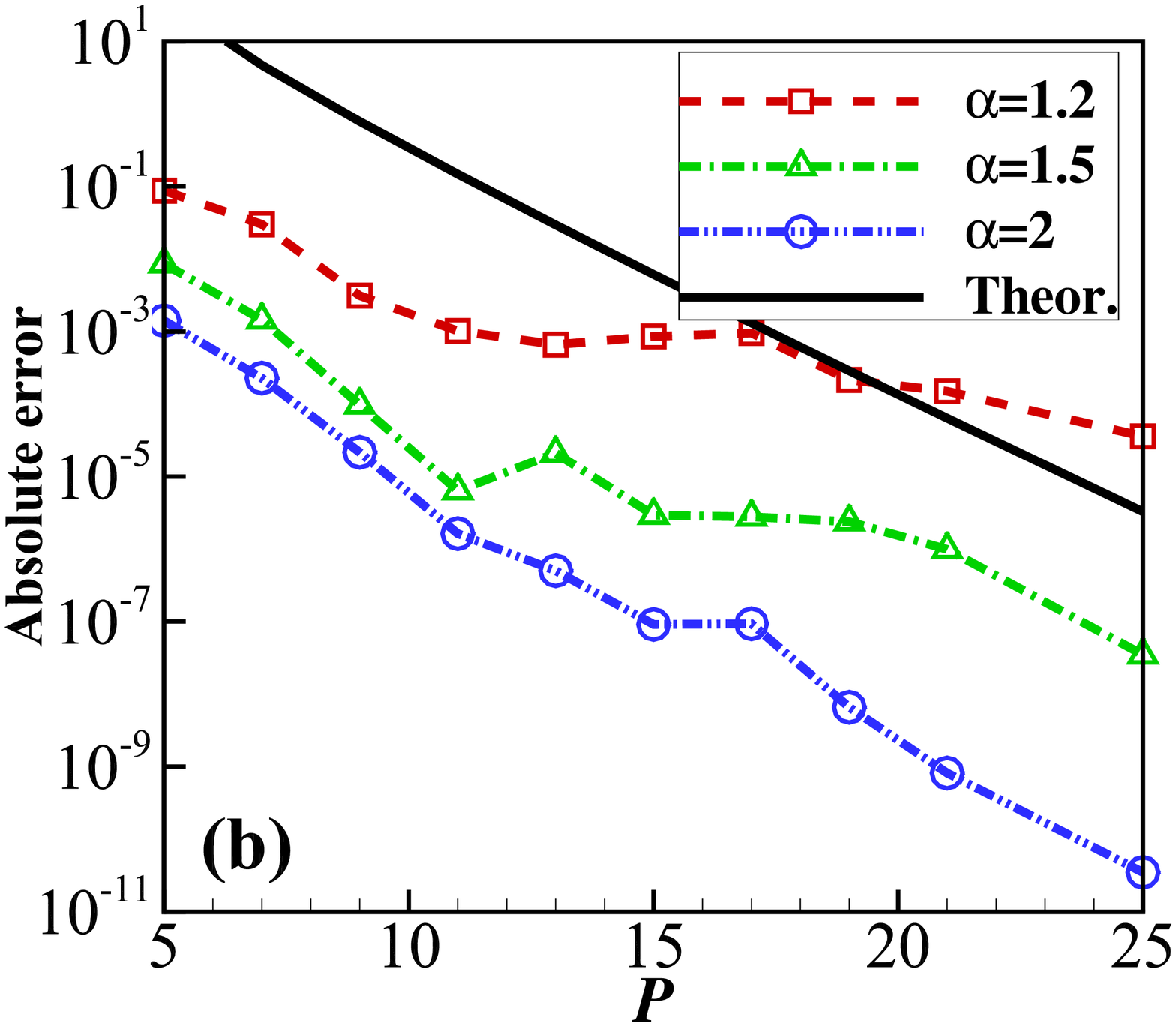}
	\includegraphics[width=0.49\linewidth]{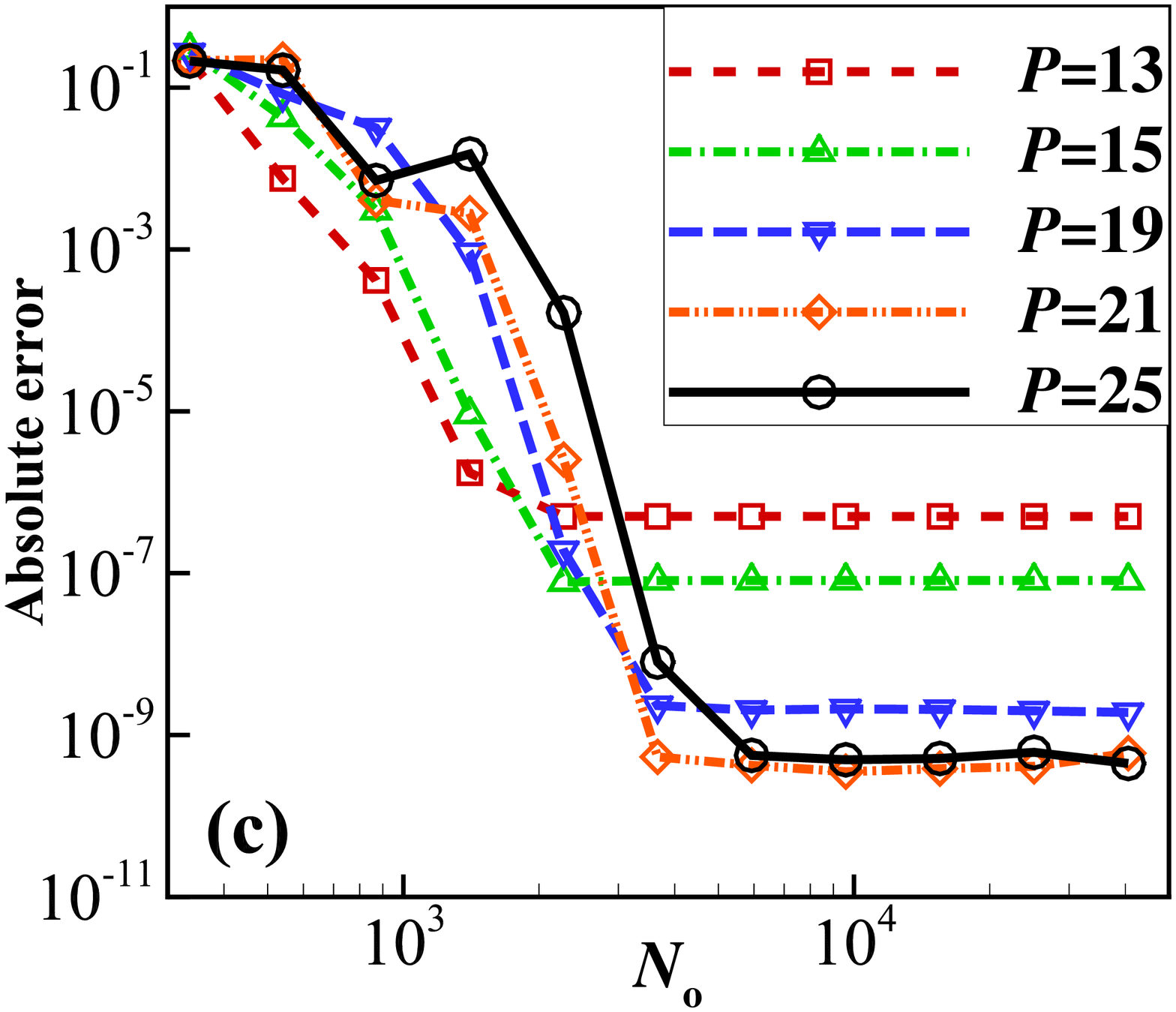}
	\includegraphics[width=0.49\linewidth]{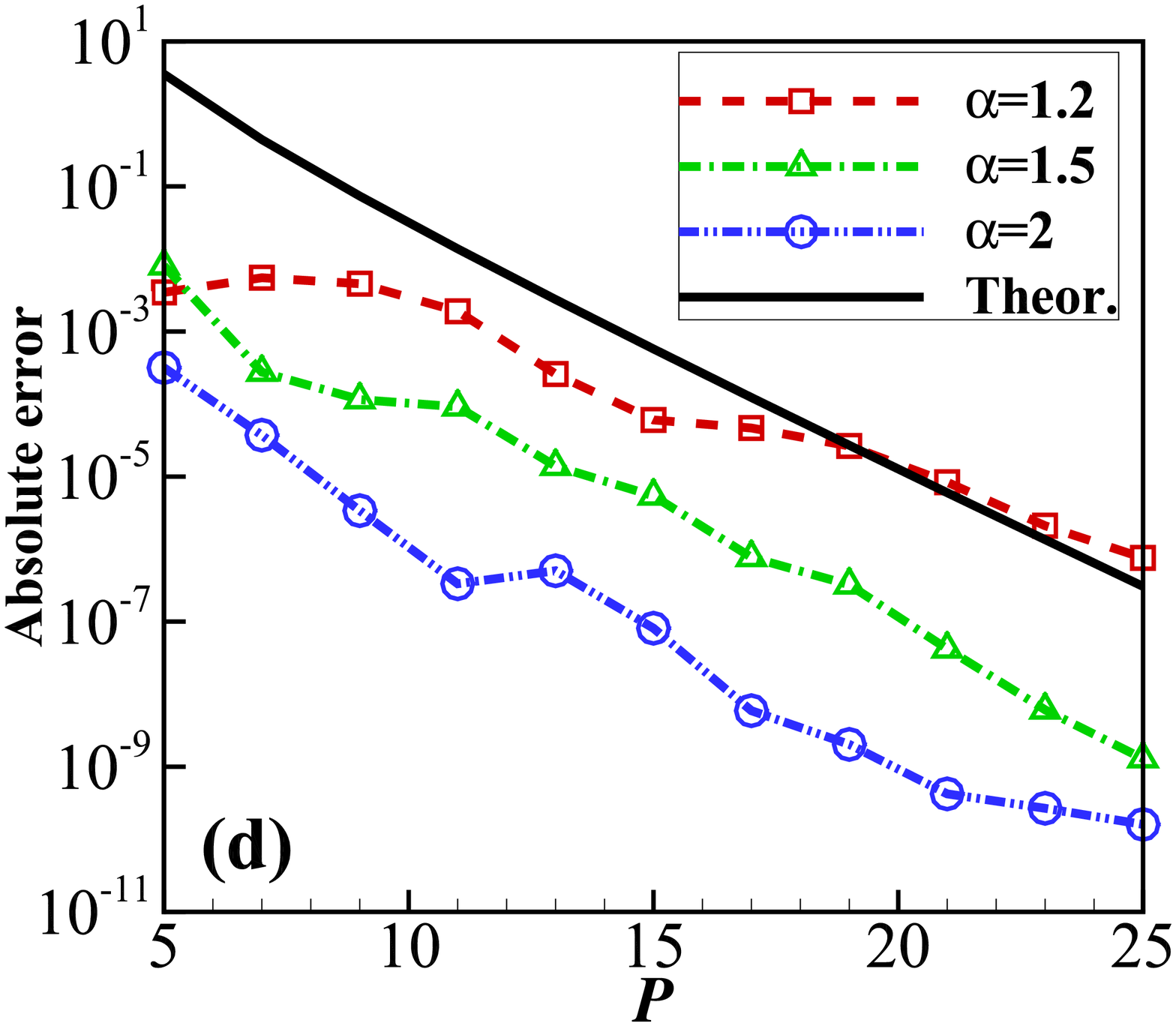}
	\caption{Errors in the potential of the HSMA as functions of the number of the integration points for different $P$ and
the number of the spherical harmonic bases for different $\alpha$. The theoretical lines
correspond to parameter $\alpha=2$. (ab) Periodic boundary condition; (cd) Dirichlet
boundary condition.}
	\label{64error}
\end{figure*}

\begin{figure*}[t!]
	\centering
	\includegraphics[width=0.49\linewidth]{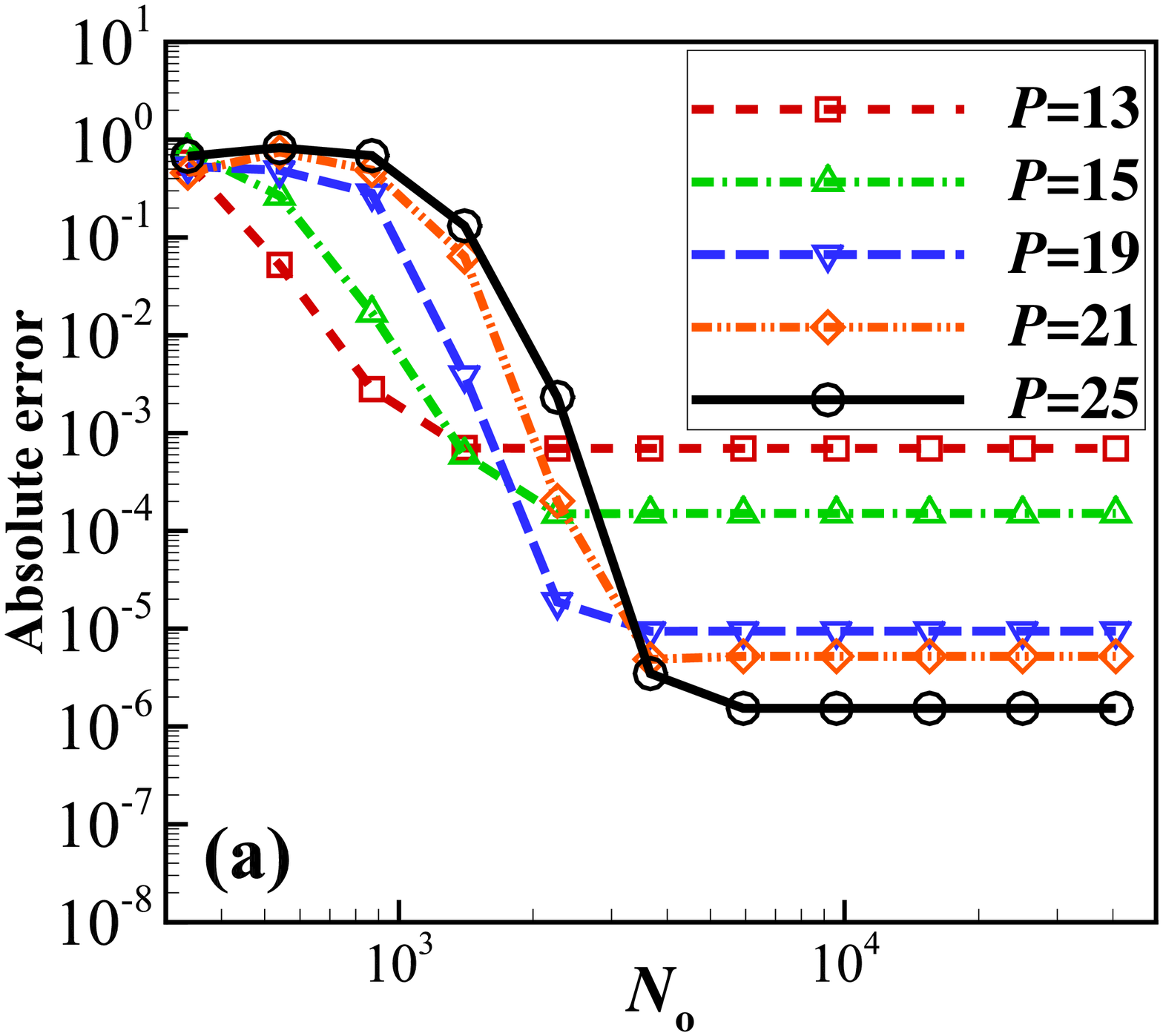}
	\includegraphics[width=0.49\linewidth]{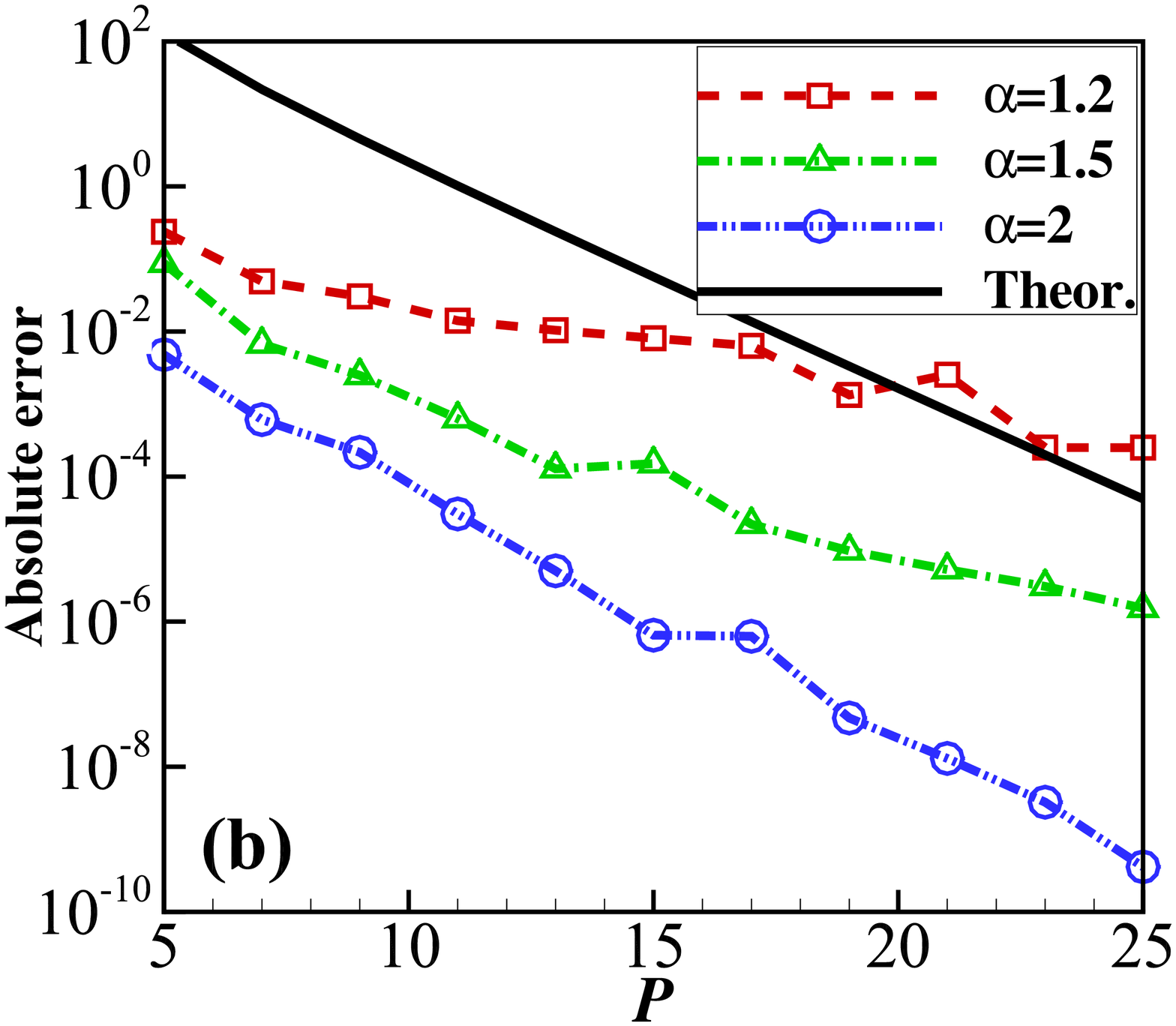}
	\includegraphics[width=0.49\linewidth]{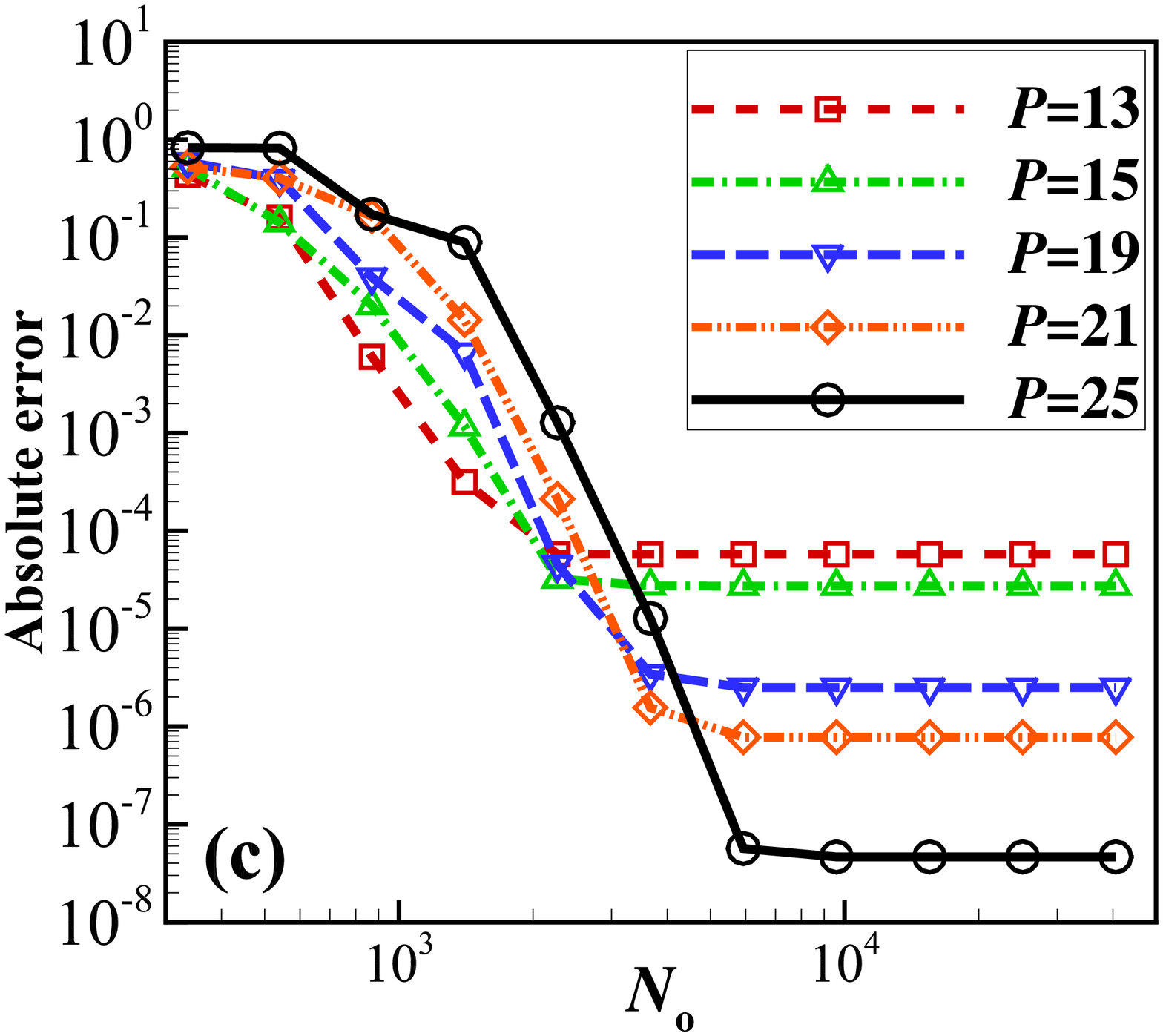}
	\includegraphics[width=0.49\linewidth]{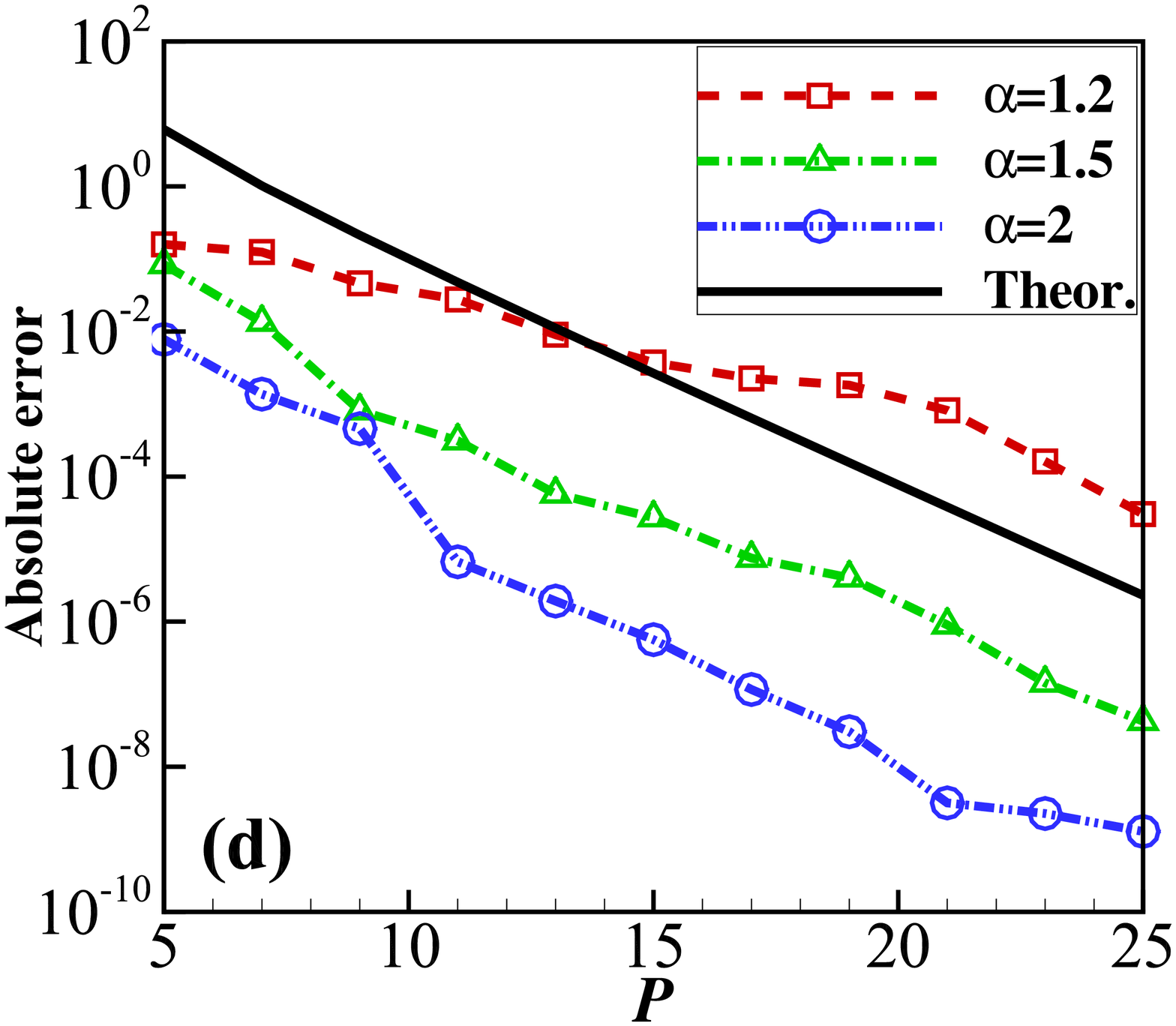}
	\caption{Errors in the force of the HSMA as functions of the number of the integration points for different $P$ and
the number of the spherical harmonic bases for different $\alpha$. The theoretical lines
correspond to parameter $\alpha=2$. (ab) Periodic boundary condition; (cd) Dirichlet
boundary condition.}
	\label{forceError}
\end{figure*}

The performance of the HSMA is tested by three examples. The computer program of the algorithm can be found on GitHub \cite{hsmaWeb}. In the calculations, the central box is cubic.
The radius of $\partial\Omega_\mathrm{s}$ takes $R_\mathrm{s}=\sqrt{3} L\alpha/2$ where $L$ is the edge length of the central box
and $\alpha>1$ is a parameter to determine the size of the auxiliary surface. The number of monitoring points is fixed to be
$N_\mathrm{m}=2P^2$ which approximates $N_\mathrm{b}$. The grid size for the central difference takes $\Delta r=10^{-5}R_\mathrm{s}$
if without additional description.

In the first example, the Madelung constant of a NaCl cubic-like crystalline lattice is calculated.
The Madelung constant is used to determine the electrostatic energy of an ion in a crystal.
In the setup, $64$ unit source charges evenly arranged on a lattice grid of edge length $L=2$, with neighboring charges
having opposite charge amount. The PBC is specified. The numerical error is computed by comparing the exact value of the Madelung constant
$-1.74756459463318219$ \cite{madelung1918elektrische}. Fig. \ref{64error}(a) shows the error convergence with
the number of integration points $N_\mathrm{o}$, where {$\alpha=2$ is taken} and the results of five different $P$
are displayed. All the curves reach platforms when $N_\mathrm{o}>3600$, i.e., the Fibonacci number in the
numerical integration $F_2\geq 610.$ The platform in the panel means that the dominate error source is from other parts
of approximation, demonstrating the rapid convergence of the numerical integration.
Fig. \ref{64error}(b) shows the error convergence with the increase of the truncated degree of the spherical harmonic bases $P$,
where $N_\mathrm{o}=5928$, i.e., the Fibonacci number $F2=987$, is fixed and the results of three different $\alpha$ are illustrated, together
with the theoretical estimate with $\alpha=2$ ($d_0$ takes the average number density in the central box). It is observed that the rate of convergence is in agreement with
the theoretical estimate and the actual error is much smaller than that of the estimate because the charge neutrality
leads to the error cancellation.

In the second example, we take $L=2$ as before but specify a Dirichlet boundary condition (DBC) on the boundary of
the central box which includes three source points: a $+2$ charge at $(0.1,0,0)$ and
two $-1$ charges at $(0.8,0.8,0.8)$ and $(-0.9,-0.9,-0.9)$ in Cartesian coordinates. The same accuracy test
as the first example is made and the results are shown in Fig. \ref{64error}(cd). In the results, the
error in the total electrostatic energy is calculated, where the reference ``exact" solution is obtained by
the HSMA method but setting $\alpha=10$ and $P=30$. The image series given by Eq. \eqref{diel} at the conducting limit is divergent
and we take $\Delta r=10^{-4}R_\mathrm{s}$ to avoid the influence of roundoff error to the accuracy. Similar
performance as the PBC can be observed from the two panels.

In molecular dynamics simulations, the gradient of the potential is calculated to obtain
the force of each charged particle. It is important to validate the accuracy of the force calculation of the
HSMA. We use the same systems with the PBC and the DBC as the previous two examples
and calculate the forces on each charge for varying parameters.
We calculate the maximum absolute value of three force components of all the source charges, and compare it
with the ``exact" reference solution for the error. The reference solution is calculated using the force balance condition
for the PBC and using the HSMA with high-accurate parameters ($\alpha=10$ and $P=30$) for the DBC.
The results are present in Fig. \ref{forceError}. Similar performance on the error convergence
as the potential calculation can be observed except that the error values are about an order of
magnitude higher than those of the potential. Fig. \ref{forceError}(bd) also shows the agreement on
the convergence rate between the numerical error and the theoretical bound in the case of $\alpha=2$.

\begin{figure*}[t!]
	\centering
	\includegraphics[width=0.49\linewidth]{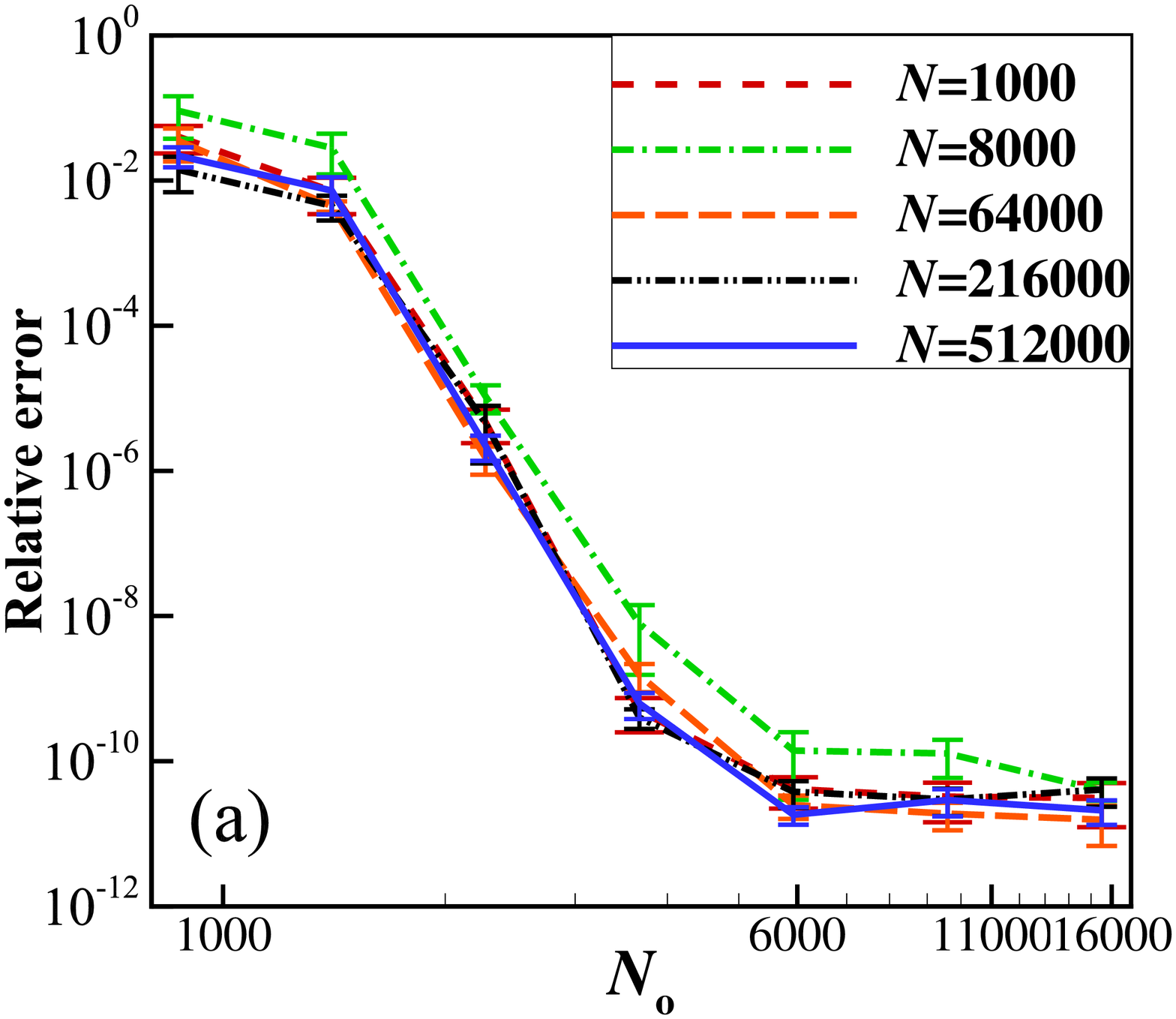}
	\includegraphics[width=0.49\linewidth]{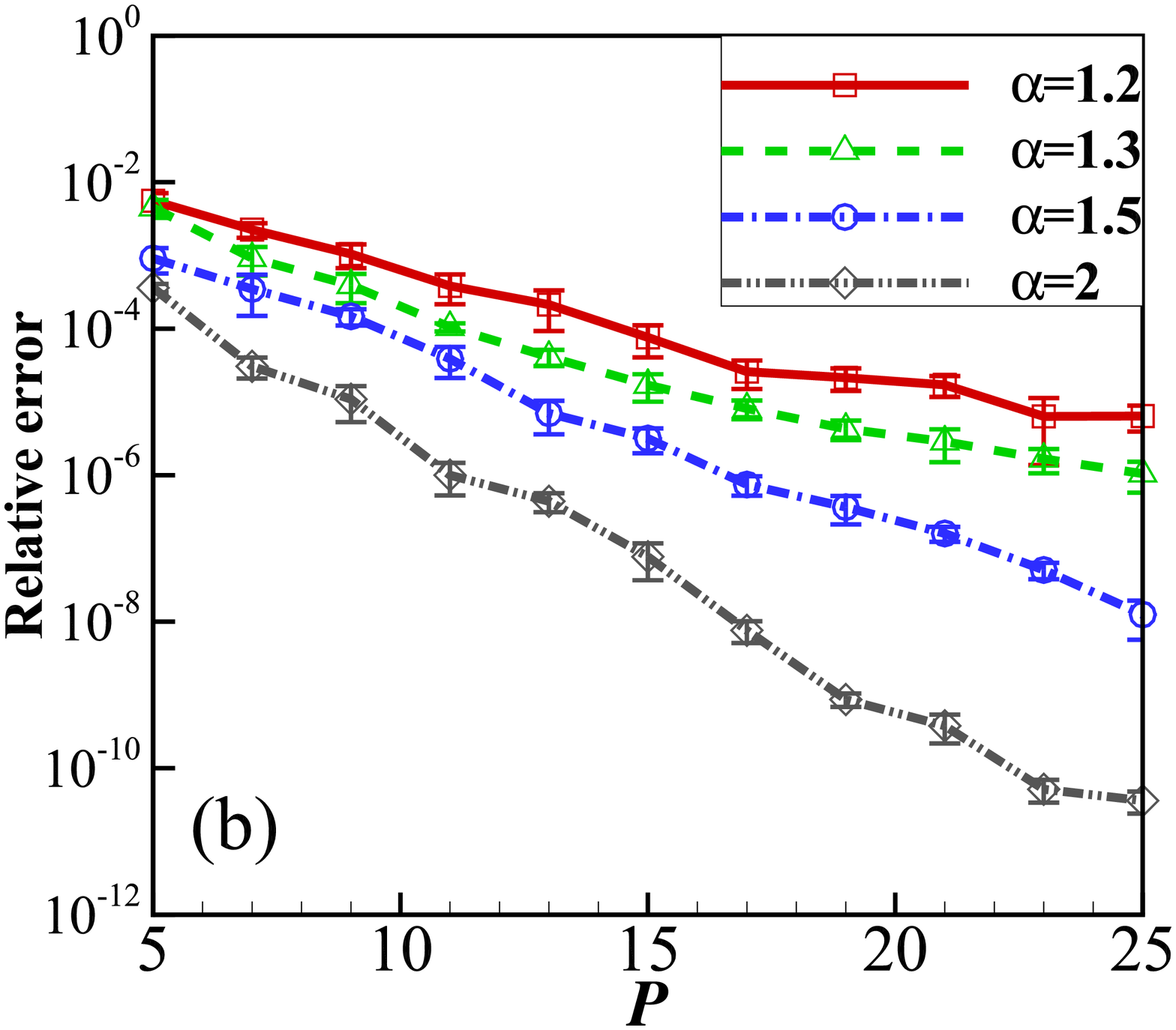}
	\caption{Errors of the HSMA for the potential as function of (a) the number of the integration points for different $N$,
and (b) the number of the spherical harmonic bases for different $\alpha$ and given $N=512k$.}
	\label{Big_error}
\end{figure*}

Next consider systems with the increase of the number of source charges $N$ in the central box.
These charges are initially distributed at the sites of a uniformly spaced lattice,
each with unit charge of random sign, and the charge neutrality is remained during the initialization.
The PBC is used and the energy by the Ewald summation method \cite{Ewald1921Die} is calculated as the ``exact" reference solution.
The GPU acceleration for the pairwise summation is used for solving the linear-squares problems and
the energy calculation.
Fig. \ref{Big_error}(a) shows the error convergence as function of the integration points for systems from
$1k$ to $512k$ particles using parameters $\alpha=2$ and $P=25$. The error bars represent the standard deviation from mean errors
for 5 runs of different initial distributions. These curves has almost the same tendency, in accord with those
present in Fig. \ref{64error} (a), demonstrating that the convergence is almost independent
of the number of source points. Fig. \ref{Big_error}(a) displays the results as function of $P$ with
varying $\alpha$ for given $N=512k$ and $N_\mathrm{o}=5928$. Again, the error decreases rapidly with
the increase of $P$ and using a larger $\alpha$ will significantly improve the results, which is
in agreement with the results in Fig. \ref{64error}(b) and the theoretical prediction.

\begin{figure}[t!]
		\centering
		\includegraphics[width=0.98\linewidth]{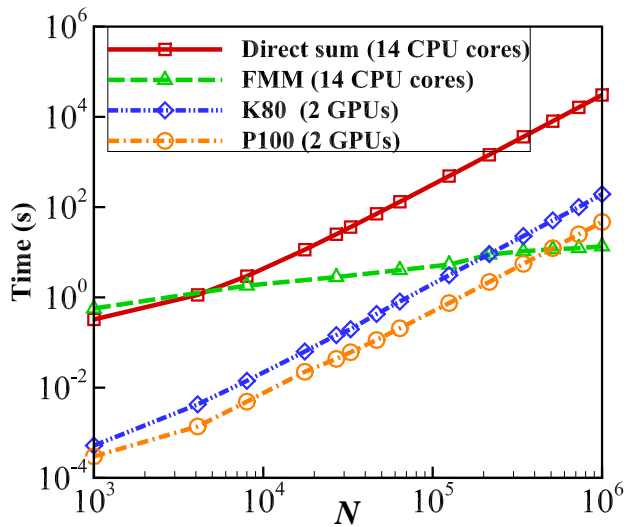}
\caption{Comparison of the computation times for the HSMA accelerated with multiple cores, FMM and GPUs (both K80 and P100).
The parameters are $\alpha=1.3, P=25$ and $N_\mathrm{o}=5928$. }
		\label{totaltime}
	\end{figure}

We now move to the timing of the HSMA for these systems. Both the FMM and the GPU accelerations are used together
with the direct sum. The simulations of the direct sum and the FMM acceleration are run on an Intel Xeon
E5-2680 v4(14 Cores, 2.40GHz, 35MB Cache, 9.6GT) machine, and all cores are used for each calculation.
The publicly available software package FMM3DLIB \cite{CGR:JCP:1999,GR:AN:1997} is adopted for the FMM acceleration, where
the FMM precision is set as $10^{-6}$. The GPU calculations are run using two NVIDIA Tesla P100 GPUs or two NVIDIA Tesla K80 GPUs.
In the GPU acceleration, we utilize $N_\mathrm{s}$ threads divided into several blocks to
calculate pairwise interactions. If the number of particles is large,
in order to reduce the shared-memory-bank conflicts in the evaluation, one optimizes the storage structure
of particle descriptions or uses the GPU whose architecture supports concurrent reads from multiple threads
to a single shared memory address \cite{Cuda2007}. We set the HSMA parameters $\alpha=1.3, P=25$ and $N_\mathrm{o}=5928$. 	
Fig. \ref{totaltime} illustrates the timing results with the increase of the source charge number. Both the brute force summations
with multicore and GPU accelerations show a quadratic scaling while the FMM acceleration shows a linear scaling.
By comparing the FMM with the direct summation, the breakeven point is $\sim 4k$ for the HSMA use.
When the source number $N$ is less than $100k$, the
brute calculation with the GPU acceleration shows promising and an improvement of $3-4$ orders of magnitude in the
time cost can be observed by comparing it with the direct summation. It is remarked
that the performance of different methods depends on how the algorithm is implemented, and the breakeven point
can be largely varied if the FMM or the direct summation is optimized. In practice, the most important is the
simulation time used for each step. We can observe from Fig. \ref{totaltime} that both the GPU times are less
than 1 second for $N=100k$, and the FMM time is slightly over 1 second, showing that the HSMA will be promising
for practical molecular dynamics simulations.

\section{Concluding remarks}

In summary, the HSMA is proposed for evaluating the electrostatic sum in a cubic box with a general boundary condition.
The algorithm transforms the electrostatic problem into the pairwise summation of finite charges, which can be easily accelerated
by the FMM to achieve a linear complexity or by the GPUs to reach a 3-4 orders of improvement in time cost.
One important feature of the HSMA is its high accuracy. For the mapped integrals, errors from the numerical integration and the approximation
of the surface dipole using the central difference is small by increasing the integration points with
minor influence to the time cost.  The error mainly comes from the truncation of the spherical harmonic series since a small number
of terms $P$ is expected for the practical use. Overall, the HSMA can be considered as a method of spectral convergence.
For a large-scale system, $P$ can be slightly increased to reduce the radius of the auxiliary surface, and thus the time cost can be significantly reduced.

The HSMA can be very useful in many simulations of molecular systems, in particular, when the system is nonperiodic
and the Ewald-based lattice summation is not very efficient, e.g., when a solid boundary is present near an electrolyte. Moreover, the HSMA
can be promising to serve as a fast solver of the Poisson's equation with a general source term, where the source can be
discretized as the sum of many point sources and thus the continuation within the auxiliary surface can be performed.

In many applications, partial periodic systems with dielectric jumps are often studied, for example, electrolytes between two
charged dielectric interfaces \cite{AH:APS:2005}. It is worthy to note that the minimization step for the harmonic expansion may be not straightforward
when the boundary is a dielectric interface as the interface conditions require the solution information of the exterior
domain of the simulation box in the form of an infinite pairwise Coulomb sum \cite{YLL:JPCB:2002,XCC:CCP:2011}. In this case,
the representation of the harmonic series expansion with to-be-determined coefficients is
also required, leading to a bigger linear system from the minimization step. The HSMA approach for this problem is our
ongoing project. The comparison of the HSMA with the Ewald-based algorithms, the implementation for molecular dynamics simulations, and more applications
of the HSMA are also the future work.

\section*{Acknowledgements}
The authors acknowledge the financial support from the Natural Science Foundation of China (Grant Nos: 11571236 and 21773165)
and the support from the HPC center of Shanghai Jiao Tong University, and thank Mr. Yichao Wang
from the HPC center for the discussion on the GPU implementation. The authors also thank the anonymous reviewers for their useful
comments and suggestions.

\appendix
\section{Error bounds of the potential gradient}

We estimate the error of the potential gradient using the truncated spherical harmonic series \eqref{harmonic}
for the approximation of $\Phi_\mathrm{out}$ which is due to all images outside the auxiliary sphere $\Omega_\mathrm{s}$.

Let $R_0$ be the radius of the circumsphere of the central box and $\alpha=R_\mathrm{s}/R_0>1.$
For any charge $q$ at $\mathbf{r}_i$ outside the auxiliary sphere, the potential can be written as,
\begin{equation}
\phi(\mathbf{r})=\frac{q}{|\mathbf{r}-\mathbf{r}_i|}=\sum_{n=0}^\infty \frac{q r^n}{r_i^{n+1}}P_n(\cos\varphi),
\end{equation}
where $\varphi$ is the angle of $\mathbf{r}$ and $\mathbf{r}_i.$ If one truncates the series at $n=P$,
and defines the truncation error of its gradient by $\mathbf{E}=(E_1, E_2, E_3)^T$, then,
\begin{equation}
\mathbf{E}=\left|\nabla\sum_{n=P+1}^\infty \frac{q r^n}{r_i^{n+1}}P_n(\cos\varphi)\right|.
\end{equation}
By a simple calculation, one can find that each component of the error vector satisfies,
\begin{equation}\
E_\ell\leq \sum_{n=P+1}^\infty \frac{n r^{n-1}}{r_i^{n+1}}=\left(\frac{r}{r_i}\right)^P \frac{(r_i-r)(P+1)+r}{r_i(r_j-r)^2}.
\end{equation}

Now for all images outside $\Omega_\mathrm{s}$, the number density of the image particles
can be written as $n(\mathbf{r})=\sum_{\mathrm{r}_i\in \Omega_\mathrm{s}^c} \delta(\mathbf{r}-\mathbf{r}_i)$.
Let $\epsilon_\mathrm{grad}^\ell$ be the $\ell$th component
of the three-dimensional error vector of the potential gradient using the
truncated spherical harmonic expansion \eqref{harmonic}. Let $q_\mathrm{m}=\max|q_i|$. Then we have,
\begin{equation}
\begin{split}
\epsilon_\mathrm{grad}^\ell&\leq \sum_{\mathrm{r}_i\in \Omega_\mathrm{s}^c}  |q_i|\left(\frac{r}{r_i}\right)^P \frac{(r_i-r)(P+1)+r}{r_i(r_i-r)^2} \\
&\leq q_\mathrm{m} \int_{\Omega_\mathrm{s}^c} n(\mathbf{x})
\left(\frac{R_0}{x}\right)^P \frac{C}{x} d\mathbf{x}  \\
&\leq 4\pi d_0 q_\mathrm{m} C \int_{R_\mathrm{s}}^\infty \left(\frac{R_0}{x}\right)^P x dx,
\end{split}
\end{equation}
where $C=[(R_\mathrm{s}-R_0)(P+1)+R_0]/(R_\mathrm{s}-R_0)^2$, and we have used the properties $r<R_0$ and $r_i>R_\mathrm{s}$ in the second inequality.
$d_0$ is a constant which approximates the average number density $n_0=N/V$ with
$V$ being the volume of the central box, $d_0\approx n_0$. An accurate value of $d_0$
depends on the uniformity of the charges, and the upper bound is an approximation if we replace
it by $n_0$.  Calculating the integral and using the
relation $\alpha=R_\mathrm{s}/R_0$, we finally obtain an error bound,
\begin{equation}
\epsilon_\mathrm{grad}^\ell\leq\frac{4\pi d_0R_s q_\mathrm{m}}{(\alpha-1)(P-2)}\left(P+\frac{\alpha}{\alpha-1}\right)\left(\frac{1}{\alpha}\right)^{P-1},
\end{equation}
which shows the spectral convergence with the order of spherical harmonic expansion.
The error bound linearly depends on the number density of the particle, and
the ratio $\alpha$ plays the most important role for the accuracy of the approximation.


\end{document}